\documentclass[prl, twocolumn,superscriptaddress,nobibnotes]{revtex4}
\usepackage{graphicx}
\usepackage{amssymb}
\usepackage{amsmath}
\usepackage{bm}
\usepackage{color}
\usepackage{float}
\usepackage{setspace}
\usepackage{physics}
\usepackage{subfigure}

\large

\begin{document}

	\title{Single-shot spatial instability and electric control of polariton condensates \\ at room temperature}
	
	\author{Ying Gao}
	\affiliation{Department of Physics, School of Science, Tianjin University, Tianjin 300072, China} 
	\affiliation{Institute of Molecular Plus, Tianjin University, Tianjin 300072, China}
	
	\author{Xuekai Ma}
	\affiliation{Department of Physics and Center for Optoelectronics and Photonics Paderborn (CeOPP), Universit\"{a}t Paderborn, Warburger Strasse 100, 33098 Paderborn, Germany}
	\affiliation{Guangdong Provincial Key Laboratory of Nanophotonic Functional Materials and Devices, South China Normal University, Guangzhou 510631, China}
	
	\author{Xiaokun Zhai}
	\affiliation{Department of Physics, School of Science, Tianjin University, Tianjin 300072, China} 
	\affiliation{Institute of Molecular Plus, Tianjin University, Tianjin 300072, China}
	
	\author{Chunzi Xing}
	\affiliation{Tianjin Key Laboratory of Low Dimensional Materials Physics and Preparing Technology, School of Science, Tianjin University, Tianjin 300072, China}
	
	\author{Meini Gao}
	\affiliation{Tianjin Key Laboratory of Low Dimensional Materials Physics and Preparing Technology, School of Science, Tianjin University, Tianjin 300072, China}
	
	\author{Haitao Dai}
	\affiliation{Tianjin Key Laboratory of Low Dimensional Materials Physics and Preparing Technology, School of Science, Tianjin University, Tianjin 300072, China}
	
	\author{Hao Wu}
	\affiliation{ Lab of quantum detection and awareness, Space Engineering University Beijing 101416, China} 
	
	\author{Tong Liu}
	\affiliation{ Lab of quantum detection and awareness, Space Engineering University Beijing 101416, China} 
	
	\author{Yuan Ren}
	\affiliation{ Lab of quantum detection and awareness, Space Engineering University Beijing 101416, China} 
	
	\author{Xiao Wang}
	\affiliation{College of Materials Science and Engineering, Hunan University, Changsha 410082, China}
	
	\author{Anlian Pan}
	\affiliation{College of Materials Science and Engineering, Hunan University, Changsha 410082, China}
	
	\author{Wei Hu}
	\affiliation{Guangdong Provincial Key Laboratory of Nanophotonic Functional Materials and Devices, South China Normal University, Guangzhou 510631, China}
	
	\author{Stefan Schumacher}
	\affiliation{Department of Physics and Center for Optoelectronics and Photonics Paderborn (CeOPP), Universit\"{a}t Paderborn, Warburger Strasse 100, 33098 Paderborn, Germany}
	\affiliation{Wyant College of Optical Sciences, University of Arizona, Tucson, AZ 85721, USA}
	
	\author{Tingge Gao}
	\affiliation{Department of Physics, School of Science, Tianjin University, Tianjin 300072, China} 
	\affiliation{Institute of Molecular Plus, Tianjin University, Tianjin 300072, China}
	
	\begin{abstract}

		In planar microcavities, the transverse-electric and transverse-magnetic (TE-TM) mode splitting of cavity photons arises due to their different penetration into the Bragg mirrors and can result in optical spin-orbit coupling (SOC). In this work, we find that in a liquid crystal (LC) microcavity filled with perovskite microplates, the pronounced TE-TM splitting gives rise to a strong SOC that leads to the spatial instability of microcavity polariton condensates under single-shot excitation. Spatially varying hole burning and mode competition occurs between polarization components leading to different condensate profiles from shot to shot. The single-shot polariton condensates become stable when the SOC vanishes as the TE and TM modes are spectrally well separated from each other, which can be achieved by application of an electric field to our LC microcavity with electrically tunable anisotropy. Our findings are well reproduced and traced back to their physical origin by our detailed numerical simulations. With the electrical manipulation our work reveals how the shot-to-shot spatial instability of spatial polariton profiles can be engineered in anisotropic microcavities at room temperature, which will benefit the development of stable polariton-based optoeletronic and light-emitting devices.
	\end{abstract}
	
	\maketitle
	
	
	Mode instability has been reported in different systems resulting from various physical properties, such as the transverse mode instability in laser physics \cite{fibre laser1, fibre laser2}, modulational instability in nonlinear optics~\cite{nonlinear book} and atomic condensates~\cite{atom soliton}, as well as dynamical instability in hybrid exciton-polariton condensates~\cite{dynamic instability}. Exciton polaritons are created when strong coupling between excitons and cavity photon modes in microcavities occurs. Being a composite Boson, below the Mott density exciton polaritons can experience a similar Bosonic condensation  as cold atoms \cite{polariton BEC1, polariton BEC2}, and the polariton condensation can be observed at much higher temperature even up to room temperature \cite{room temperature GaN, organic polariton4, organic polariton5}. The polariton lifetime is very short in the picosecond range such that the exciton polariton is a non-equilibrium system. Under non-resonant pumping, an exciton reservoir can be excited and acts as a source for the polariton condensate; the interaction between the exciton reservoir and the condensate plays a critical role in the condensation process\cite{polariton exciton reservoir interaction 1, polariton exciton reservoir interaction 2, polariton exciton reservoir interaction 3}. The non-equilibrium nature and the nonlinear interaction between the polaritons and the exciton reservoir can lead to the instability of polariton condensates, forming phase defects~\cite{instability theory1, instability theory3} or filamentation patterns~\cite{dynamic instability,  GaAs single shot}. For example, in organic microcavities, the polariton condensate shows shot-to-shot fluctuation under a large pumping spot, while it transforms into nearly uniform spatial distribution when the excitation spot size is decreased \cite{dynamic instability}. In GaAs based microcavities, the polariton condensate under single-shot excitations shows that its stability against filamentaion is very sensitive to the energy relaxation rate and the detuning, which are closely related to the polariton and exciton reservoir interaction  \cite{GaAs single shot}. 
	
	It is known that the TE-TM splitting can appear within microcavities due to the different penetration of TE and TM modes into the Bragg mirrors that are used to confine different cavity photon modes \cite{TE-TM1, TE-TM2}. In the spinor formalism, the TE-TM splitting acts as an effective magnetic field, which influences the distribution of different spin polarized polaritons and results in the optical spin Hall effect \cite{kavokin spin hall effect, Non Abelian TETM} and the directional flow of polaritons or periodic oscillation of the pseudospin \cite{spin hall effect nature physics, nonlinear optical spin hall effect, TETM vortex}. Under large uniform pumping, the spin-orbit coupling (SOC) induced by the TE-TM splitting leads to unbalanced interaction with the exciton reservoir for different spin components, thus the polariton condensate distribution in the near field can be modified significantly. The previous works regarding the instability of the polariton condensates only focus on the polaritons condensing in a single branch, that is, the intrinsic TE-TM splitting as well as the polarization properties are not explicitly considered. How the polariton stability is affected by the TE-TM splitting induced SOC  remains unexplored. Typically, TE-TM splitting in a given sample is an intrinsic property and cannot be easily manipulated during the measurements.

	Liquid crystal (LC) molecule based microcavities offer a platform to electrically control the linearly polarized cavity photon modes~\cite{1-liquid crystal_science} as well as the polariton modes~\cite{Yao RD}. The voltage applied to the microcavity rotates the LC molecule's director orientation, thus the horizontally linearly polarized polariton modes can be tuned whereas the vertically linearly polarized polariton modes are unaffected. The electrical manner to tune the linear polarizations enables the manipulation of the two modes and the resulting SOC. In other words, when the TE and TM modes are near resonant, their energy splitting can cause an effective magnetic field for the generation of the SOC. When the two linear modes are away from each other by applying the voltage onto the microcavity the effective magnetic field vanishes. A tunable optical spin Hall effect has been reported in a LC based optical microcavity~\cite{liquid crystal Light}.

	Here, by electrically controlling the TE-TM splitting, we systematically analyze how the TE-TM splitting influences the stability of the polariton condensate. The stable polariton condensate is observed at 0 V, where the two linear polarizations are completely separated without SOC and the condensate only occupies one of them. At 2.8 V, the polariton condensate is transformed to a strong shot-to-shot variation structure originating from the SOC under intense femtosecond laser excitation, where the two linear polarizations are brought into resonance and both components are nearly equally occupied. Under a higher voltage of 4.1 V, the two linearly polarized modes are lifted again, such that the polaritons condense only at the vertically linearly polarized branch. As a consequence, the polariton condensate from each single-shot excitation becomes spatially stable again. The experimental observations are well reproduced by our numerical simulations. Our work paves the way to investigate the tunability of the polariton stability against filamentation at room temperature as well as to study the underlying physics of the non-equilibrium state transition between unstable to stable ones.
	
	\begin{figure}
		\centering
		\includegraphics[width=\linewidth]{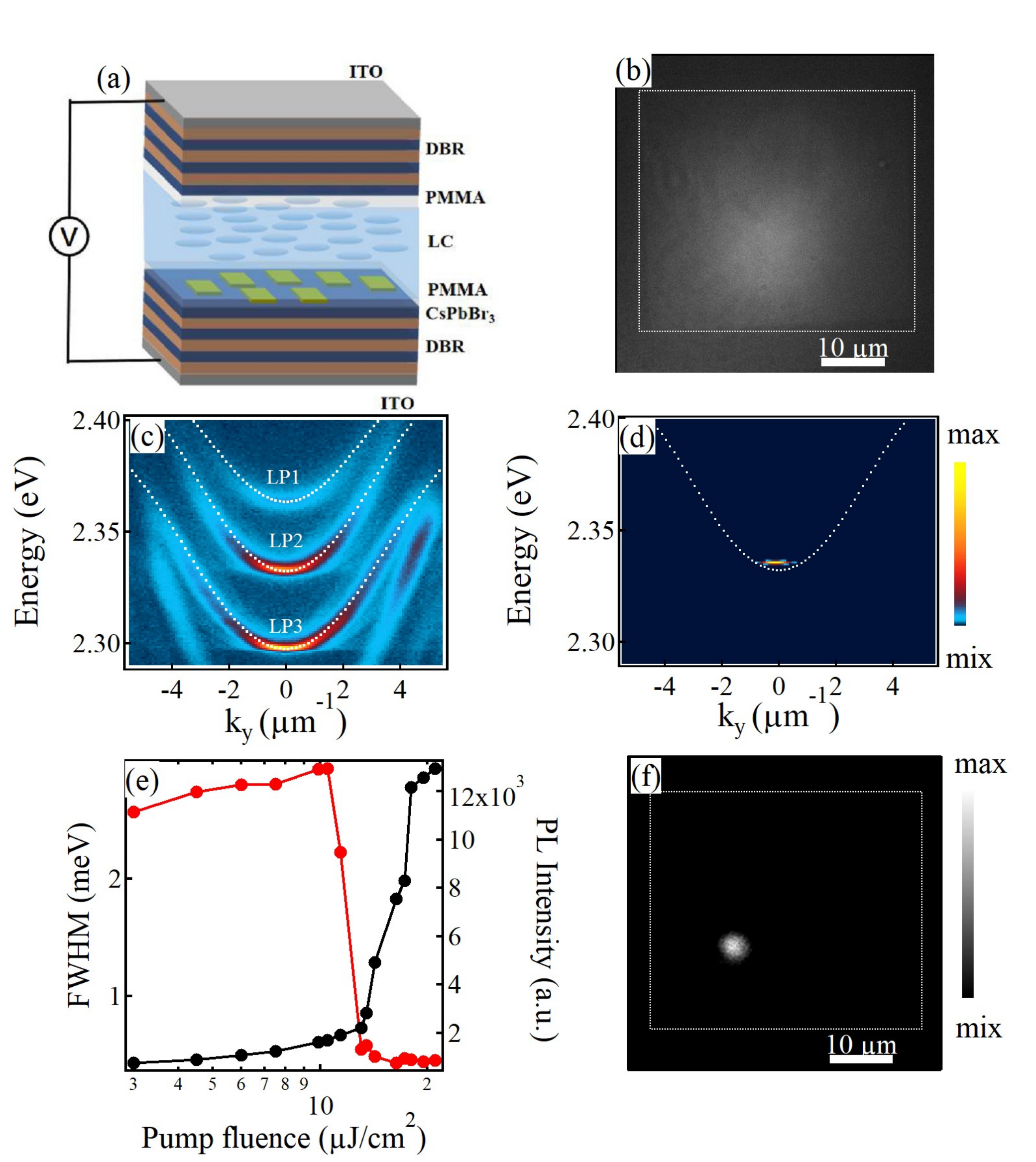}
		\caption{\textbf{Exciton polariton condensation in a liquid crystal (LC) microcavity.} (a) The microcavity filled with a LC as the cavity spacer layer and with CsPbBr$_3$ microplates inserted as the gain material.  (b) The optical imaging of the perovskite microplate. (c) Photoluminescence dispersion of the microcavity pumped by a femtosecond laser below the threshold.The detuning of LP1, LP2 and LP3 are -28 meV, -70 meV and -100 meV respectively. (d) Photoluminescence dispersion of the microcavity pumped by a femtosecond laser above the threshold. The dash line in (c) and (d) are the fitted polariton dispersion. (e) Integrated intensity and linewidth of the polariton against the pumping density. (f) Near field imaging of the polariton condensate under the voltage of 0 V. }
	\end{figure}

	\begin{figure}
		\centering
		\includegraphics[width=\linewidth]{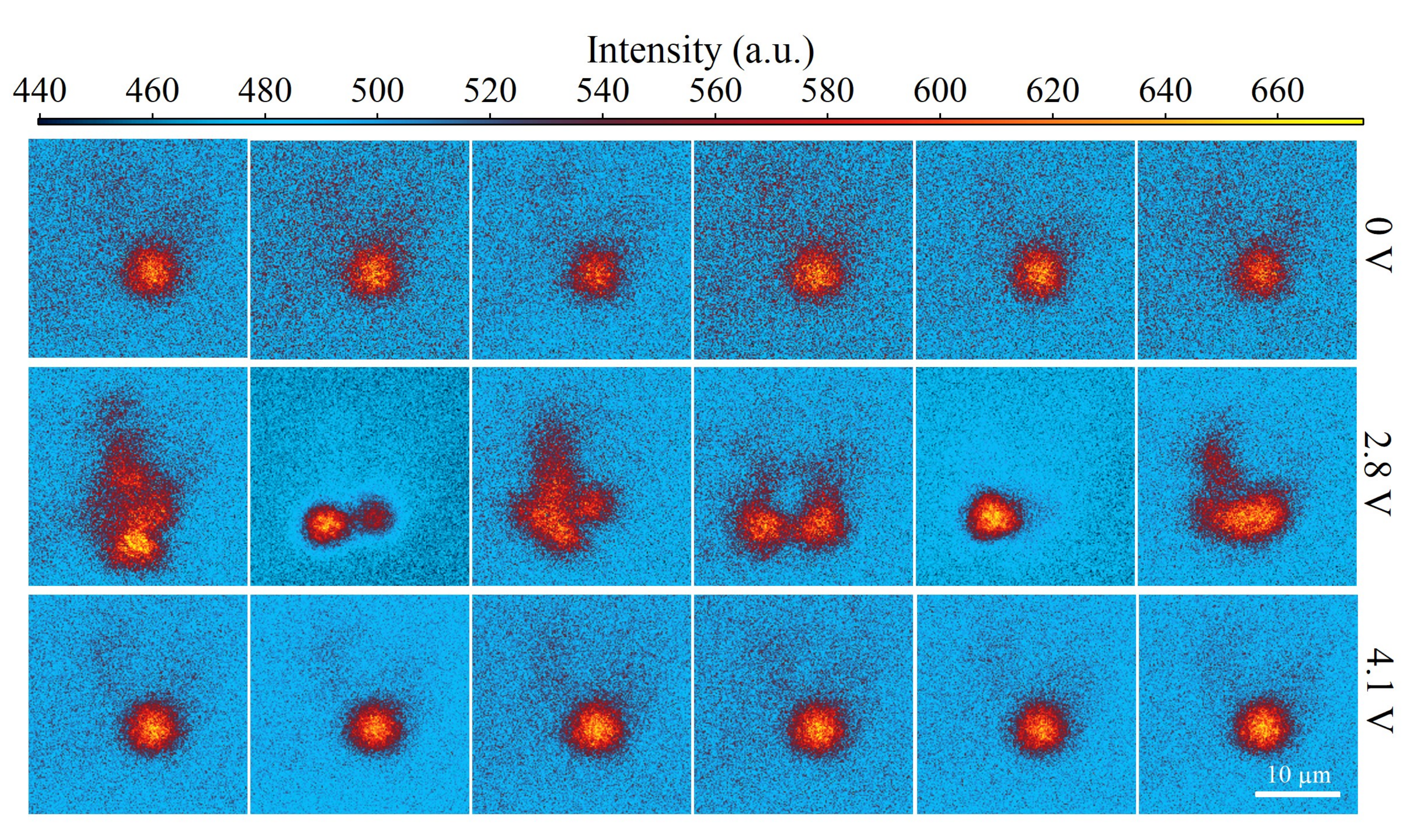}
		\caption{\textbf{Single-shot near field distribution of the polariton condensate under different voltages.} The top panel, middle panel, and bottom panel correspond to the shot-to-shot spatial polariton condensate distribution taken under the voltage of 0 V, 2.8 V, and 4.1 V, respectively. In each panel, 6 continuous near field imaging under 6 laser excitation taken by the camera are shown. In the top and bottom panels, stable near field imaging of polariton condensates are visible, however, a strong fluctuating polariton distribution can be seen in the middle panels. The pumping density is fixed at 1.5 $P_\textup{th}$.} 
	\end{figure}
	
	
	In the experiments, the CsPbBr$_3$ microplates are grown using the CVD method on a mica substrate. Then the perovskite microplates are transferred onto the bottom DBR (SiO$_2$/Ta$_2$O$_5$) with the center wavelength of the stop band at 530 nm. The ordering layers (PMMA) with the thickness of around 200 nm is deposited onto the perovskite microplates and the top DBR. The microcavity is formed by pasting the two DBRs, and finally the LC molecules are filled into the microcavity (Fig. 1(a)). 
	
	The microcavity dispersion is measured using a home-made angle-resolved spectroscopy with a femtosecond laser as the pumping source (LightConversion system, 5700 Hz, 400 nm, spot size 40 $\mu$m, horizontally linear polarized). Fig. 1(b) shows the near field image of the microplate that has been chosen in this work. As shown in Fig. 1(c), several polariton modes are observed because of the larger cavity length which supports multiple cavity modes. The strong coupling between the excitons in the CsPbBr$_3$ microplate and the multiple cavity modes creates several lower polariton branches LP1-LP3 (other branches are not shown since they do not contribute to the results in this work). The detailed fitting parameters of the polariton modes LP1-LP3 and corresponding exciton and photon components using the coupled oscillator model are shown in the SM. Under 0 V, the LC molecule is aligned along \textit{x} direction by standard rubbing procedure. The applied electric field rotates the LC molecule director, so that the horizontally linearly polarized modes LP1 and LP3 can be tuned by increasing the voltage onto the microcavity, whereas LP2 is vertically linearly polarized ones which are immune to the electric field. Under the pumping density of around 14 $\mu$J/cm$^2$, the integrated intensity of the emitted photons from the microcavity under the voltage of 0 V shows obvious superlinear increase against the pumping density, and the linewidth of the polariton mode is greatly reduced, evidencing the occurrence of the polariton condensation at the ground state of the LP2 [see Fig. 1(d,e)]. In the near field, we observe a Gaussian-shape profile of the condensate [Fig. 1(f)] located in a local area of the microplates due to the inhomogeneity, i.e., disorder, of the sample.

	The single-shot experiments can reveal the fluctuation of the polariton condensate in the near field and the condensation process \cite{interaction strength}. In the following we perform the single-shot experiments by using a delay generator which synchronizes the detecting camera and the pumping laser \cite{GaAs single shot}. The experimental results with increasing the voltage from 0 V to 4.1 V are shown in Fig. 2. Under 0 V, a stable pattern of the polariton condensate is observed in every single laser pulse excitation (captured by 6 continuous realizations, of the polariton condensate are plotted in the top panel of Fig. 2). The variation of the polariton intensity can be neglected for different excitation experiments. Under 0 V, the polaritons mainly condense at LP2 with vertical linear polarization, whereas the population of the horizontally linearly polarized component LP3 can be neglected (see Fig. 3(a, b)). 

	Increasing the voltage to 2.8 V, the near field imaging of the polariton condensate shows strong shot-to-shot fluctuation, as shown in the middle panels of Fig. 2. The shot-to-shot fluctuation of the polariton condensate can be observed between the pumping density of 1.5 $P_\textup{th}$ and 2 $P_\textup{th}$ as the electric field remains unchanged. Similar condensate filamentation have been observed in \cite{GaAs single shot, dynamic instability}. Especially in GaAs-based microcavities~\cite {GaAs single shot}, the filamentation of the polariton condensate is determined by the detuning between the excitons and cavity photon modes. When the detuning is negative and the relaxation of the high energy excitons is inefficient, the polariton condensate shows random shot-to-shot fluctuation, which disappears under positive detuning where the polariton-reservoir interaction enhances the condensation process. In our experiments, from Figure 3(c, d), one can see that LP3 is tuned to be resonant with LP2 to create a prominent TE-TM splitting (2.1 meV at 2 $ \mu$m$^{-1}$, which is much larger than GaAs microcavities, see the SM), in virtue of the modulation of the XY splitting by the electric field. The large TE-TM splitting under the voltage of 2.8 V influences the distribution of polaritons with different spins via the SOC and thereby the interaction with the exciton reservoir. The resulting unbalance interaction with the exciton reservoir of the two spin components acts as to reshape the condensates differently in each single pulse excitation on set of the condensation process and leads to the random fluctuation of the near field imaging demonstrated by the middle panels of Fig. 2.

	\begin{figure}
		\centering
		\includegraphics[width=\linewidth]{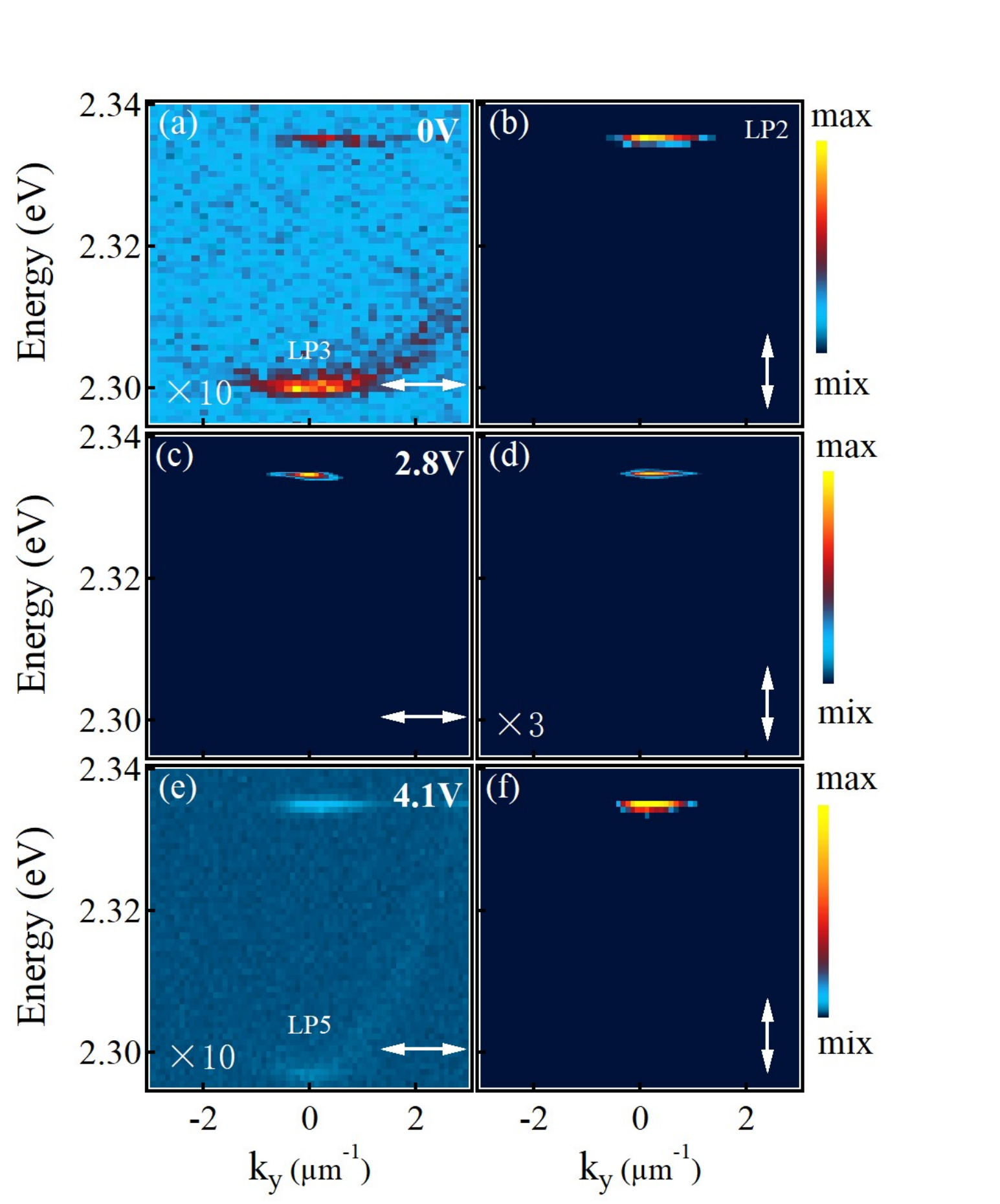}
		\caption{\textbf{Linearly polarized dispersion of the polariton condensate.} Horizontally and vertically linearly polarized dispersion taken at 0 V (a, b), 2.8 V (c, d), and 4.1 V (c, d). The arrows indicate the polarization directions. (a) and (e) is multiplied by 10, (d) is multiplied by 3 for the guide of eyes. The horizontal linear polarized modes dominate under the voltage of 0 and 4.1 V.}
	\end{figure}
	\setlength{\floatsep}{5pt plus 2pt minus 2pt}
	\setlength{\textfloatsep}{5pt plus 2pt minus 2pt}
	\setlength{\intextsep}{5pt plus 2pt minus 2pt}

	Different from GaAs microcavities where the single-shot fluctuation can only be tuned by moving the sample to the area where the detuning is more positive, our work offers a simple method to tune the near field distribution of polariton condensate. In our experiment, when the voltage is larger than 2.8 V, the mode LP3 are tuned to the higher energy, and as a result the polaritons condense mainly to LP2 and show vertical linear polarization [Figure 3(e, f)] under the pumping density of 1.5 $P_\textup{th}$. In this case, the TE-TM splitting of the polariton condensate can be ignored. From the the near field measurements, one can see that the polariton condensate is stable up to a stronger voltage 4.1 V (see the bottom panels of Fig. 2). Generally speaking, as the applied voltage increases from 0 V to 4.1 V, the polariton condensate experiences a stable-unstable-stable gradual process. The transition states at 2.5 V and 3.4 V can be found in the SM. The sensitivity of the polariton instability against the voltage in the experiments clearly shows the role of the TE-TM splitting within the microcavity.
	

	\setlength{\parskip}{0.2cm plus4mm minus3mm}
	
	To investigate the principle of the instability observed in our experiments, we mimic the dynamics of polariton condensates by using the binary Gross-Pitaevskii model, i.e., in the circular polarization basis~\cite{nonlinear optical spin hall effect}, 
	\begin{equation}
	\begin{aligned}\label{eq:GP}
	i\hbar\frac{\partial\Psi_{\pm}(\mathbf{r},t)}{\partial t}=&\left[-\frac{\hbar^2}{2m}\nabla_{\bot}^{2}-i\hbar\frac{\gamma_\text{c}}{2}+g_\text{c}|\Psi_{\pm}(\mathbf{r},t)|^2 \right. \\
	&\left.+\left(g_\text{r}+i\hbar\frac{R}{2}\right)n_{\pm}(\mathbf{r},t)+V(\mathbf{r})\right]\Psi_{\pm}(\mathbf{r},t) \\
	&+\frac{\Delta_\text{LT}}{k_\text{LT}^2}\left(i\frac{\partial}{\partial x}\pm\frac{\partial}{\partial y}\right)^2\Psi_{\mp}(\mathbf{r},t). \\
	\end{aligned}
	\end{equation}
	Here $\Psi_{\pm}(\mathbf{r},t)$ is the wavefunction of the polariton condensate and the subscripts $\pm$ denote the spin components. $m$ is the effective mass of the polariton condensate. $\gamma_{c}$ is the loss rate in quasi-mode approximation \cite{Carcamo:20} which can be compensated by the gain from the exciton reservoir $n_{\pm}(\mathbf{r},t)$ with a condensation rate $R$. $g_\text{c}$ and $g_\text{r}$ are the polariton-polariton interaction and polariton-reservoir interaction, respectively. $V(\mathbf{r}$) is the disorder potential with the correlation length $\sim$ 5 $\mu$m and the depth $\sim$ 0.5 meV, representing the rugged sample surfaces, to approximate the experimental condition. $\Delta_\text{LT}$ is the TE-TM splitting at a finite momentum $k_\text{LT}$. It is worth pointing out that in our case we assume that the incoherent exciton reservoir $n_{\pm}(\mathbf{r},t)$ is reshaped by both spin components, because during the condensation process the phase and polarization information of the "hot" excitations are not preserved. Therefore, the equation of motion for the exciton reservoir satisfies
	\begin{equation}
	\begin{aligned}\label{eq:reservoir}
	\frac{\partial n_{\pm}(\mathbf{r},t)}{\partial t}=&\left[-\gamma_\text{r}-R\left(|\Psi_{\pm}(\mathbf{r},t)|^2+|\Psi_{\mp}(\mathbf{r},t)|^2\right)\right]n_{\pm}(\mathbf{r},t)  \\
	&+P_{\pm}(\mathbf{r},t)\,.
	\end{aligned}
	\end{equation}
	Here $n_{\pm}(\mathbf{r},t)$ is the density of the reservoir. $\gamma_\text{r}$ is the loss rate of the reservoir. $P_{\pm}(\mathbf{r},t)$ is the incoherent and pulsed pump with a standard Gaussian distribution in both real space (diameter $\sim$20 $\mu$m) and time (duration $\sim$50 ps) (see the detailed definition in the SM), non-resonantly driving the system for the creation of the polariton condensates. Here we consider a linearly polarized pump with $P_{+}=P_{-}$.

	\begin{figure}
		\centering
		\includegraphics[width=\linewidth]{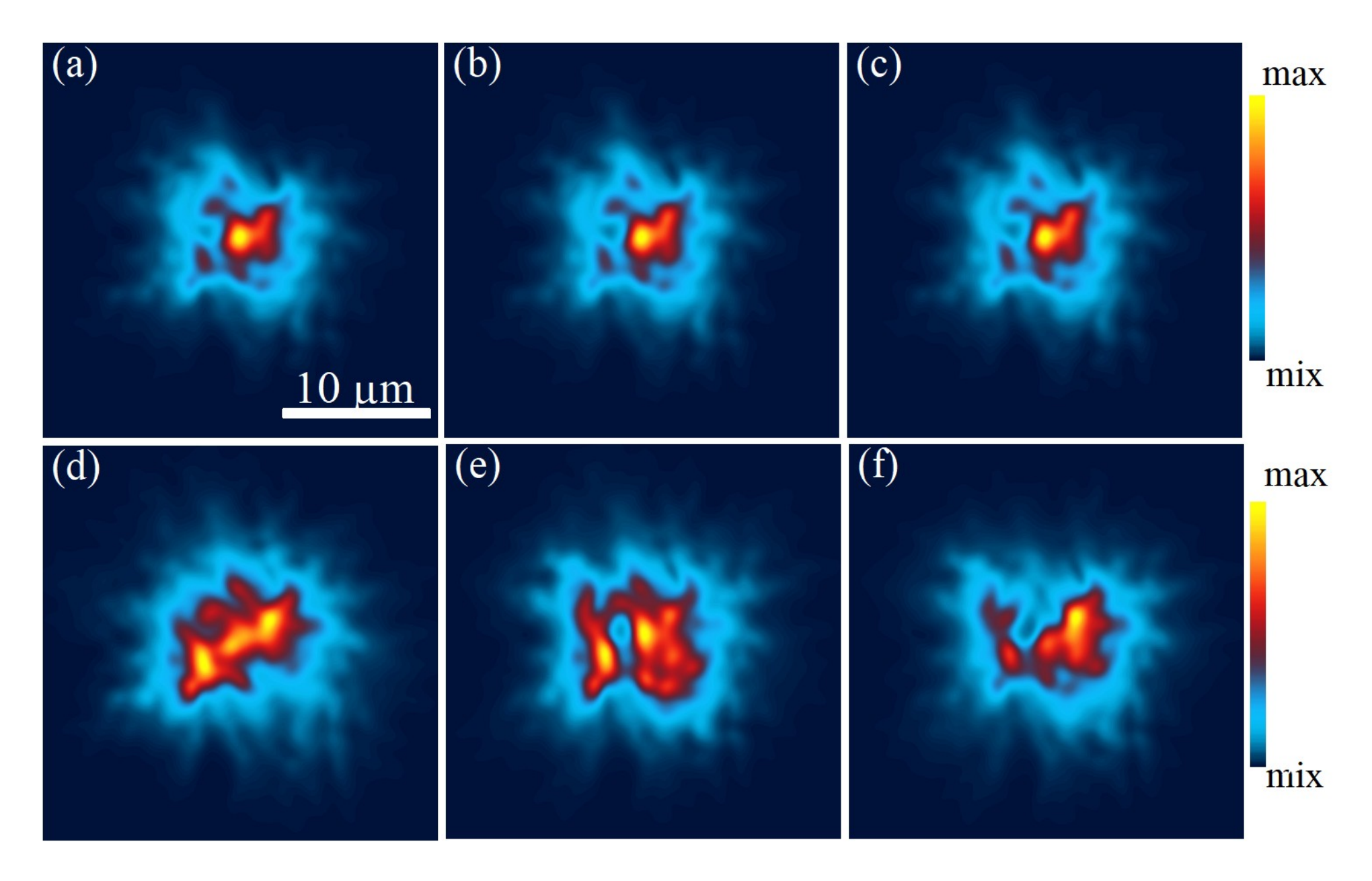}
		\caption{\textbf{Numerical results of the single-shot excitation.} Time-integrated polariton densities (linearly $x$-polarized component, i.e., $|\Psi_{+}+\Psi_{-}|^2$) for three single-shot excitations at (a-c) $\Delta_\textup{LT}=0$ in which the solutions are almost identical and (d-f) $\Delta_\textup{LT}=1$ meV where the solutions are distinct. Other parameters for numerical simulations are: $m=0.2$ $m_\textup{e}$ ($m_\textup{e}$ is the free electron mass), $\gamma_\textup{c}=0.2$ ps$^{-1}$, $\gamma_\textup{r}=1.5\gamma_\textup{c}$, $R=0.01$ ps$^{-1}$ $\mu$m$^{2}$, $g_\textup{c}=0.6$ $\mu$eV $\mu$m$^{2}$, $g_\textup{r}=2g_\textup{c}$, and $k_\textup{LT}=2$ $\mu$m$^{-1}$.}
	\end{figure}

	In the above model, the static electric field or its direct influence on the dynamics of polariton condensates is not included. However, from the experimental results, one can see that the two linearly polarized cavity modes can be brought into resonant at around 2.8 V where a clear TE-TM splitting, $\sim$2.1 meV at $k=2$ $\mu$m$^{-1}$, is measured (see the SM). This energy splitting, which vanishes when the two modes are completely separated at other voltages, gives rise to a strong SOC, corresponding to the last term on the right side of Eq. \eqref{eq:GP}. From the numerical results, one can see that without SOC, i.e., $\Delta_\text{LT}=0$, the distribution of the condensate at each independent excitation is almost identical [Fig. 4(a)-4(c)], while the situation changes when the SOC become significant as shown in Fig. 4(d)-4(f) in which each identical pump pulse creates a different density pattern (from initial noise). This is because the SOC induces different density distributions of the two spin components, such that their contributions to the reshaping of the reservoir are distinct. The reshaped reservoir enables the polaritons to condense in the adjacent local energy minima in the disorder. The numerical results are nicely consistent with the experimental observations shown in Fig. 2, demonstrating that the instability originates from the strong TE-TM splitting. The density profiles in Fig. 4 are linearly x-polarized, i.e., $|\Psi_{+}+\Psi_{-}|^2$. Every density profile obtained in the numerical simulation is integrated over the entire simulation period ($\sim$ 100 ps), and in each single-shot excitation a different initial noise condition is applied. Note that the TE-TM splitting used in the numerical simulation is 1 meV at $k_\textup{LT}=2$ $\mu$m$^{-1}$ which is smaller than that measured in experiment. The reason for the disagreement between the experiment and theory is that the linewidth of the polariton modes below the threshold measured in the experiment is broad (see the SM), such that this value may be overestimated. 
	
	In our work, one can see that the polaritons condense in the ground state of LP2 and LP3 under the voltage of 2.8 V (Figure 3(c, d)), hence the shot-to-shot fluctuation does not originate from the random transition between the discrete energy levels. This excludes the possibility of the appearance of the transverse mode instability induced long-time random fluctuation resulted from heating effect or quasi-periodic modulation of the refractive index as discussed in the fibre laser system \cite{fibre laser1} (note that the pumping power we used in our experiment is only 1.5 $P_\textup{th}$). In addition, the polariton condensate distribution can be tuned from unstable to stable without the need to change the exciton/photon components which is necessary in GaAs microcavities \cite{GaAs single shot}. Moreover, comparing with the modulational instability in nonlinear systems which may become stable when the input power is changed, the instability we observed can be engineered to be stable by simply tuning the voltage onto the microcavity, and the electric field does not directly influence the electrically neutral polariton condensates.

	To summarize, we have experimentally and theoretically studied how to electrically tune the spatial distribution stability of the polariton condensate in a LC microcavity. In such systems, the instability of the polariton condensates is due to the strong SOC which occurs when the two linearly polarized polariton modes are brought into resonance by an electric field. As the two linear modes are spectrally separated from each other, the SOC disappears, resulting in the stable polariton condensation. The electrical manipulation of the polariton condensates studied in our work offers to develop polariton-based optoelectronic devices, circuits and explore lasers with stable transverse mode profiles.

	\begin{acknowledgments}
		T.G. acknowledges the support from the National Natural Science Foundation of China (NSFC) (No. $12174285$), H.D. appreaciates the support from the NSFC (No. $61975748$), Y.R. thanks the support from the NSFC (No $62173342$), and the Paderborn group acknowledges the support from the Deutsche Forschungsgemeinschaft (DFG) (Grant No. 467358803 and No. 519608013).
	\end{acknowledgments}

	

\end{document}